\documentstyle[aps]{revtex}
\begin{document}
\title{Crossover from thermal hopping to quantum tunneling in Mn$_{12}$Ac}
\author{S. P. Kou$^{1,2}$, J. Q. Liang$^{1,2}$,Y.B. Zhang$^{1.2}$, X.B.Wang$^3$ and
F. C. Pu$^{1,4}$}
\address{$^1$Institute of Physics and Center for Condensed Matter \\
Physics, Chinese\\
Academy of Sciences, Beijing 100080, P. R. China\\
$^2$Department of Physics, Shanxi University, Shanxi 030006, P. R. China\\
$^3$Center for Advanced Study, Tsinghua University, Beijing 100084, P.\\
R.China\\
$^4$Department of Physics, Guangzhou Normal College, Guangzhou 510400,\\
P.R.China.}
\maketitle

\begin{abstract}
The crossover from thermal hopping to quantum tunneling is studied. We show
that the decay rate $\Gamma $ with dissipation can accurately be determined
near the crossover temperature. Besides considering the
Wentzel-Kramers-Brillouin (WKB) exponent, we also calculate contribution of
the fluctuation modes around the saddle point and give an extended account
of a previous study of crossover region. We deal with two dangerous
fluctuation modes whose contribution can't be calculated by the steepest
descent method and show that higher order couplings between the two
dangerous modes need to be taken into considerations. At last the crossover
from thermal hopping to quantum tunneling in the molecular magnet Mn$_{12}$%
Ac is studied.
\end{abstract}

PACS number(s): 75.45.+j, 75.50.Tt.

\section{Introduction}

The decay of metastable states in macroscopic systems is a fundamental
problem in many areas of physics, such as macroscopic quantum tunneling in
Josephson system\cite{1,2}, violation of baryon-lepton in Weinberg-Salam
model, nucleation in first order phase transition theory\cite{3,14} and more
recent magnetic quantum resonant tunneling\cite{4}. The crossover from
thermal hoping to quantum tunneling has been studied intensively. Using
functional integral approach, Affleck first demonstrated the transition can
be found between classical regime and quantum regime\cite{6}. Larkin and
Ovchimikov also suggested it and gave a formula determining the boundary of
first order transition and second order one\cite{8,9}. Grabert and Weiss
discussed the phase transition in the presence of dissipative effects of the
environment in some detail\cite{10,11,12}.

At high temperature, the decay of the metastable state is determined by
process of thermal activation, which is governed by the Arrhenius law, $%
\frac{\omega _0}{2\pi }\exp (-\Delta U/T)$ where $\omega _0=\sqrt{-U^{\prime
\prime }(x_0)/M}$ is the well frequency and $\Delta U$ is the barrier
height. While at $T=0$ the particle can escape from the metastable state due
to quantum tunneling, the rate of which goes as $\exp (-B)$ where $B$ is the
WKB exponent. Ignoring the prefactor and equating the exponents, one obtains
the estimate 
\begin{equation}
T_0^{(0)}=\frac{\Delta U}B,  \label{1.1}
\end{equation}
where the superscript at $T_0$ means that the ground state tunneling is
considered. For $T>T_0^{(0)},$ one has practically $\Gamma \cong \Gamma
_{therm}(T),$ whereas below the transition $\Gamma \cong \Gamma _{quan}$ is
independent of temperature.

It turns out that for common metastable or double-well potentials, such as
cubic or quartic parabola, below the crossover temperature $T_0$ the
particles cross the barrier at the most favorable energy level $E(T)$ which
goes down from the top of the barrier to the bottom of $U(x)$ with lowering
temperature. The second-order transition from the classical thermal
activation to thermally assisted tunneling (TAT) is smooth and the
transition temperature is given by 
\begin{equation}
T_0^{(2)}=\frac{\omega _b}{2\pi }  \label{1.2}
\end{equation}
where $\omega _b=\sqrt{-U^{\prime \prime }(x_b)/M}$ is the barrier
frequency, and $x_b$ corresponds to the top(the saddle point) of the barrier.

In ref. (\cite{5}), Chudnovsky stressed the analogy of this kind of
transition phenomena with ordinary phase transitions and analyzed the
general conditions for both types of quantum-classical transitions. For the
second-order transition the period of oscillation $\tau (E)$ in the inverted
potential $-U(x)$ increases monotonically with energy $E$ from the top of
the barrier. If $\tau (E)$ is not monotonic, the first-order transition
occurs. The escape rate can be conveniently represented in terms of the
effective temperature defined by 
\begin{equation}
\Gamma \sim \exp [-\frac{\Delta U}{T_0^{(1)}}]=\exp [-S_{\min }/\hbar ].
\label{1.3}
\end{equation}
The actual dependence of $S_{\min }(T)$ goes along the minimum of these two
actions ( sphelaron and periodic instantons ) and the first-order transition
occurs at $T=T_0^{(1)}.$ The first derivative of $S_{\min }(T)$ is
discontinuous at $T_0^{(1)},$ providing that the crossover from the thermal
to the quantum regime is the first order transition.

The second-order transitions are common, whereas the first-order ones are
exotic and have to be specially looked for. Nevertheless, a number of
systems and processes that show first-order transitions are already known,
e.g., a SQUID with two Josephson junctions\cite{13}, false vacuum decay in
field theories\cite{14,15,16,17} and depinning of a massive string from a
linear defect\cite{18,19}. All these systems have more degrees of freedom
than just a particle, thus search for a physical system equivalent to a
particle in a potential $U(x)$ leading to the first-order transition of the
escape rate seems quite actual. Qualitatively it is clear, how $U(x)$ look
like: The potential should change slowly near the top and the bottom, but is
rather steep in the middle. In this case, as for the rectangular barrier,
tunneling just below the top of the barrier is {\em unfavorable}, the TAT
mechanism is suppressed, and the thermal activation competes with the ground
state tunneling directly, leading to the first order transition.

Quantum tunneling of the magnetization(QTM) has become a focus of interest
in physics and chemistry because it can provide a signature of quantum
mechanical behavior in macroscopic system\cite{4,5,6,20,21,22,23,27,28}. At
low enough temperature, it has been demonstrated that the vector of the
magnetization formed by a large number of spins in magnetic system can
coherently tunnel between the degenerate minima of magnetic energy.
Theoretical suggestions have led to a number of experiments which seem to
support the idea of magnetic tunneling. Since the Mn$_{12}$Ac complex
magnetic molecule provides a more suitable model for the magnetic quantum
tunneling, extensive works have been performed to demonstrate the QTM in
large spin molecules\cite{24,25,26}. On the other hand, Mn$_{12}$Ac
molecular is one of the very few examples which could exhibit the
first-order transition.

We derived a compact formula for decay rate which is valid for the entire
range of parameters of the interest in the problem of MQT. The
quantum-classical transitions of the escape rates in the dissipation systems
are investigated by the periodic instanton method. Applying the periodic
instanton method, we showed that the first-order transitions occur below the
critical external magnetic field $h_x=\frac 14$ for Mn$_{12}$Ac molecular
which is in good agreement with earlier works\cite{6,27}. The results of the
application of a previous method is developed for dealing with the
quasi-zero modes and calculating of decay rate in the crossover region which
is beyond the steepest descent method\cite{10,11,12}. In the crossover
region of second order transition $h_x>\frac 14$, we take fourth-order terms
into account to include nonlinear couplings between the modes and obtain the
universal law in this region. The point $h_x=\frac 14$ is the boundary of
first-order and second order. At this point the fourth order couplings
between modes disappear and sixth order terms must be considered. For $h_x<%
\frac 14$ before the eigenvalues of the two quasi-zero modes reach zero, the
first order transition occurs and there is no universal law in common. While 
$T_0^{(1)}$ is not far from $T_0^{(2)}$ ( such as $T_0^{(1)}=1.078T_0^{(2)}$%
for $h_x=0.1$ ), the two dangerous modes also play important roles on the
tunneling rate and need to be calculated carefully.

\section{Decay rate in the crossover region of second order transition}

The partition function can be written as a functional integral over periodic
paths where the path probability is weighted according to the Euclidean
action 
\begin{eqnarray}
S &=&\int_0^{\beta \hbar }d\tau [\frac 12M\dot{x}^2+U(x)]  \label{2.1} \\
&&+\frac 12\int_0^{\beta \hbar }d\tau \int_0^{\beta \hbar }d\tau ^{\prime
}k\left( \tau -\tau ^{\prime }\right) x\left( \tau \right) x\left( \tau
^{\prime }\right)  \nonumber
\end{eqnarray}
where $k\left( \tau \right) =\frac 1{M\beta \hbar }\sum\limits_{n=-\infty
}^\infty \xi \left( \nu _n\right) exp\left( i\nu _n\tau \right) $, $\nu
_n=2\pi n/\beta \hbar ,$ and $\xi \left( \nu _n\right) =\gamma \left( \nu
_n\right) \mid \nu _n\mid $ is related to the frequency-dependent damping
coefficient $\gamma \left( \nu _n\right) .$ $U(x)$ is a metastable potential
with a local minimum at $x=0$ and a local maximun at $x=x_b.$ We use $\omega
_R$ to denote the solution of the following equation $\omega _R^2+\omega
_R\gamma \left( \omega _R\right) =\omega _b^2$ where $\omega _b=\sqrt{%
-U^{\prime \prime }\left( x_b\right) /M}$ characterizes the width of the
parabolic top of the well. In classical limit, $\frac 1\hbar \rightarrow
\infty $ , the steepest descents method is available:

\begin{equation}
\delta S\left[ x\left( \tau \right) \right] =0,{}\qquad x\left( 0\right)
=x\left( \beta \hbar \right) .  \label{2.2}
\end{equation}
The fluctuation modes about the saddle point are expanded using $\Psi _n$, $%
x=x_c\left( \tau \right) +\sum\limits_nY_n\Psi _n$, where $Y_n$ are
fluctuation amplitudes and $\Psi _n$ are modes of the spectrum :

\begin{eqnarray}
-\ddot{\Psi}_n+U\left[ x_c\left( \tau \right) \right] \Psi _n &=&\omega
_n^2\Psi _n,  \label{2.3} \\
\Psi \left( \beta \hbar \right) &=&\Psi \left( 0\right) .  \nonumber
\end{eqnarray}

According to the metastable-decay theory quantum tunneling rate has the form 
$\Gamma =-\frac 2hImF$. Above the crossover temperature $T_c,$ the decay
process comes from the thermal activation $\Gamma =\frac 2\hbar \frac \beta {%
\beta _c}ImF$ where $\beta _c=2\pi /\omega _R$\cite{6}$.$ In ordinary case,
the one loop correction which results in a prefactor of the WKB leading
order exponential does not enhance the tunneling significantly and the
transition rate is dominated by the WKB leading order exponential. Near
transition point the imaginary part of the free energy has a common form :

\begin{equation}
I_mF=-\frac 1{2\beta }\left( \frac{\omega _0}{\omega _b}\right) \frac{\omega
_1^{\left( 0\right) 2}}\Lambda f_c\exp \left( \frac{-S_c}\hbar \right)
\label{2.4}
\end{equation}
where $S_c$ is just the WKB leading order exponent, $\omega _n^{\left(
0\right) 2}=\omega _0^2+\nu _n^2+\nu _n\gamma \left( \nu _n\right) ,$ $%
\omega _n^{\left( b\right) 2}=-\omega _b^2+\nu _n^2+\nu _n\gamma \left( \nu
_n\right) ,$ $1/\Lambda $ comes from the two quasi-zero modes which need to
be calcaulated carefully and 
\begin{eqnarray}
f_c\left[ \omega _0,\omega _b\right] &=&\prod\limits_{n=2}^\infty \left[ 
\frac{\omega _n^{\left( 0\right) }}{\omega _n^{\left( b\right) }}\right] ^2
\label{2.5} \\
&\rightarrow &_{\gamma \equiv const}\frac{\Gamma \left( 2-\frac{\lambda
_b^{+}}{\nu _1}\right) \Gamma \left( 2-\frac{\lambda _b^{-}}{\nu _1}\right) 
}{\Gamma \left( 2-\frac{\lambda _0^{+}}{\nu _1}\right) \Gamma \left( 2-\frac{%
\lambda _0^{-}}{\nu _1}\right) }  \nonumber
\end{eqnarray}
where $\lambda _b^{\pm }=-\frac \gamma 2\pm \left[ \frac{\gamma ^2}4+\omega
_b^2\right] ,$ $\lambda _0^{\pm }=-\frac \gamma 2\pm \left[ \frac{\gamma ^2}4%
-\omega _0^2\right] $ and $\omega _0=\sqrt{-U^{\prime \prime }\left(
0\right) /M}$. Years ago Grabert and Weiss discussed the transition rate in
the presence of dissipative effects of the environment in some detail\cite
{9,10,11}. They found an unstable mode besides the zero mode near transition
point and calculated it carefully\cite{7,8,9,10,11}. {\em Near phase
transition point the fluctuation modes about the saddle points include two
dangerous modes whose contribution can't be calculated by the steepest
descent method and it is necessary to consider higher order couplings
between the two dangerous modes}\cite{7,8,9,10,11}.

\subsection{Beyond steepest descent for $T>T_0^{(2)}$}

Above $T_0^{(2)},$ the decay process is dominated by the saddle point called
sphelaron $x=x_b.$ Considering the fluctuation modes around it, we have the
periodic paths near the saddle point\cite{9,10,11} 
\begin{eqnarray}
x &=&x_b+Y_0+Y_{-1}\sqrt{2}\sin \frac{2\pi }L\tau +Y_1\sqrt{2}\cos \frac{%
2\pi }L\tau +...  \label{2.6} \\
&&+Y_{-n}\sqrt{2}\sin \frac{2\pi n}L\tau +Y_n\sqrt{2}\cos \frac{2\pi n}L\tau
...\text{.}  \nonumber
\end{eqnarray}
There is a mode with negative eigenvalue $\omega _0^{\left( b\right)
2}=-\omega _b^2=-U^{\prime \prime }/M$ which is the key mode giving
contribution to the imaginary part of the free energy. From the steepest
descent method the result of the partition function of sphelaron solution is
written into 
\begin{eqnarray}
Z &=&\int\limits_{x\left( 0\right) =x\left( \beta \hbar \right) =0}D[x\left(
\tau \right) ]\exp \{%
%TCIMACRO{\tfrac{-S\left[ x\left( \tau \right) \right] }\hbar }
%BeginExpansion
{\textstyle {-S\left[ x\left( \tau \right) \right]  \over \hbar}}
%EndExpansion
\}  \label{2.7} \\
\ &=&N\int \prod\limits_ndY_nexp\{%
%TCIMACRO{\tfrac{-S\left[ Y_n\right] }\hbar }
%BeginExpansion
{\textstyle {-S\left[ Y_n\right]  \over \hbar}}
%EndExpansion
]  \nonumber \\
\ &=&\frac 1{2i}\frac 1{\sqrt{\beta \hbar \mid \omega _b^2\mid }}\frac N{%
\sqrt{\beta \hbar \sum\limits_{n\neq 0}\omega _n^{\left( b\right) 2}}}e^{%
\frac{-S_c}\hbar }  \nonumber
\end{eqnarray}
where $N=\frac{\sqrt{\omega _0^{(0)2}}\sqrt{\sum\limits_{n\neq 0}\omega
_n^{(0)2}}}{2sinh(\frac{\beta \omega _0}2)}$\cite{29}. The eigenvalues of
the two lowest positive modes are $\lambda _1=\omega _1^{(b)2}=\omega
_{-1}^{(b)2}=\nu _1^2-\omega _R^2$. Second order transition occurs when the
eigenvalue of the lowest modes is equal to zero $\lambda _1=0$ as
temperature decreases, so it is defined that: $T_c=\omega _R/2\pi $. Near
the transition point, the eigenvalue of the lowest positive modes is : 
\begin{equation}
\lambda _1=-a\varepsilon  \label{2.8}
\end{equation}
where $\varepsilon =(1-\frac T{T_c})$ and $a=\omega _b^2+\omega _R^2\left[
1+\partial \gamma \left( \omega _R\right) /\partial \omega _R\right] .$

To regularize the divergent integral we have to add terms of fourth order in
the amplitudes $Y_{\pm 1}$. After expanding the potential about the barrier
top 
\begin{equation}
U(x)=\bigtriangleup U-\frac{M\omega _b^2x^2}2+\sum_ic_ix^i  \label{2.9}
\end{equation}
where $c_i=U^{\left[ i\right] }\left( x=x_b\right) /i!$, we obtains the
action

\begin{eqnarray}
S\left[ q\right] &=&\int_0^{\beta \hbar }d\tau \left[ \frac 12\left( \dot{x}%
_c\left( \tau \right) +\sum\limits_nY_n\dot{\Psi}_n\right) ^2+V\left(
x_c\left( \tau \right) +\sum\limits_nY_n\Psi _n\right) \right]  \label{2.10}
\\
\ &=&\hbar \beta \bigtriangleup U+\frac 12\hbar \beta m\left[
\sum\limits_{n=-\infty ,n\neq \pm 1}^\infty \omega _n^{\left( b\right)
2}Y_n^2\right] +\Delta S  \nonumber
\end{eqnarray}
where

\begin{equation}
\Delta S=\beta \hbar \left[ \frac 12m\omega _1^{\left( b\right) 2}Y_1^2+%
\frac 12m\omega _{-1}^{\left( b\right) 2}Y_{-1}^2+B_4\left(
Y_1^2+Y_{-1}^2\right) ^2\right]  \label{2.11}
\end{equation}
$B_4=$ $\frac 32c_4+\frac{9c_3^2}{2M\omega _b^2}-\frac{9c_3^2}{4M\omega
_2^{\left( b\right) 2}}$ and $\omega _2^{\left( b\right) 2}=4\nu ^2-\omega
_R^2\simeq 3\omega _R^2.$ After integrating the amplitudes $Y_0$ and $Y_n$,
we deal with the quasi-zero modes to considerate the fourth term : $%
B_4\left( Y_1^2+Y_{-1}^2\right) ^2$. Introducing the polar coordination : $%
\rho \cos \theta =Y_1,$ $\rho \sin \theta =Y_{-1},$ we get 
\begin{eqnarray}
\frac 1\Lambda &=&\frac \beta {2\pi }\int dY_1dY_{-1}\exp -\left( \beta
\Delta S\right)  \label{2.12} \\
\ &=&\frac{\kappa \sqrt{\pi }}{2\omega _R^2}%
%TCIMACRO{\limfunc{erfc} }
%BeginExpansion
\mathop{\rm erfc}
%EndExpansion
\left( -\kappa \varepsilon \right) \exp \left( \kappa ^2\varepsilon ^2\right)
\nonumber
\end{eqnarray}
where $\kappa =\frac{M\omega _R^2}2\sqrt{\frac \beta {B_4}}$. $B_4=${\em \ }$%
\frac 32c_4+\frac{9c_3^2}{2M\omega _b^2}-\frac{9c_3^2}{4M\omega _2^{\left(
b\right) 2}}=0${\em \ has been defined as the boundary between the first
order transition and the second order one} in ref. (\cite{7}). For $B_4<0$
the integration in eq. (\ref{2.12}) is divergent and this kind of divergence
will be discussed in first order transition cases. It is obvious that the
dissipation may change the boundary between the first order transition and
the second order one. Then we have the transition rate from eq. (\ref{2.4})
and eq. (\ref{2.5}) 
\begin{equation}
\Gamma =\frac{\omega _0}{2\pi }\omega _1^{(0)2}\frac{\kappa \sqrt{\pi }}{%
2\omega _R^2}%
%TCIMACRO{\limfunc{erfc} }
%BeginExpansion
\mathop{\rm erfc}
%EndExpansion
\left( -\kappa \varepsilon \right) \exp \left( \kappa ^2\varepsilon
^2\right) f_c\left[ \omega _0,\omega _b\right] e^{-U/k_BT}.  \label{2.13}
\end{equation}
Away from the crossover region the result goes back to the classical decay
rate $\frac{\omega _0}{2\pi }e^{-U/k_BT}$.

\subsection{Beyond steepest descent for $T<T_0^{(2)}$}

Below the crossover temperature, the saddle point is named by periodic
instanton or thermon. There are also two dangerous modes about this saddle
point near $T_0^{(2)}$ : one is quasi-zero mode which is associated with a
phase fluctuations of the periodic instantons with the eigenvalue and
eigenstates of $\omega _2^{\left( b\right) 2}=2a\varepsilon $ and $\Psi _2=%
\sqrt{2}\sin \left( \omega _b\tau \right) $; the other represents amplitudes
fluctuation and gives large contribution to partition function with $\omega
_3^{\left( b\right) 2}=0$ and $\Psi _3=\sqrt{2}\cos \left( \omega _b\tau
\right) $. The quasi-zero mode just takes place of ''soft mode'' which
restores symmetry and the zero mode of Goldstone mode which reflects the
freedom of phase. This is just the character of Global $U(1)$ symmetry
broken.

Near $T_0^{(2)},$ this kind of classical periodic trajectory of thermon may
be written as a Fourier series 
\begin{equation}
x_c(\tau )=\sum\limits_{n=0}^\infty \left[ X_n\cos \left( \nu _n\tau \right)
+X_{-n}\sin \left( \nu _n\tau \right) \right]  \label{2.14}
\end{equation}
The periodic paths near the saddle point are similar to that of eq. (\ref
{2.6}) $x=x_c\left( \tau \right) +\sum\limits_nY_n\Psi
_n=\sum\limits_n\left( Y_n+X_n\right) \Psi _n$. We define the amplitudes
into another form $Y_n^{\prime }=Y_n+X_n$ and the action is

\begin{equation}
S\left[ q\right] =\hbar \beta \bigtriangleup U+\frac 12\hbar \beta m\left[
\sum\limits_{n=-\infty ,n\neq \pm 1}^\infty \omega _n^2Y_n^{\prime 2}\right]
+\Delta S  \label{2.15}
\end{equation}
where 
\begin{eqnarray}
\Delta S &=&\frac 12\beta \hbar m\omega _1^{\left( b\right) 2}\left(
Y_1^{\prime }\right) ^2+\frac 12\beta \hbar m\omega _{-1}^{\left( b\right)
2}\left( Y_{-1}^{\prime }\right) ^2  \label{2.16} \\
&&\ \ +B_4\left[ \left( Y_1^{\prime }\right) ^2+\left( Y_{-1}^{\prime
}\right) ^2\right] ^2.  \nonumber
\end{eqnarray}
In term of $Y_{\pm 1}^{\prime }$ we obtain the tunneling rate below $%
T_0^{(2)}$%
\begin{equation}
\Gamma =\frac 1{\hbar \beta }\frac{\omega _0}{\omega _b}\frac{\omega
_1^{(0)2}}\Lambda f_c\left[ \omega _0,\omega _b\right] e^{-\bigtriangleup
U/k_BT}  \label{2.17}
\end{equation}
where $\frac 1\Lambda =\frac{\kappa \sqrt{\pi }}{2\omega _R^2}%
%TCIMACRO{\limfunc{erfc}}
%BeginExpansion
\mathop{\rm erfc}
%EndExpansion
\left( -\kappa \varepsilon \right) \exp \left( \kappa ^2\varepsilon
^2\right) $ and $\kappa =\frac{M\omega _R^2}2\sqrt{\frac \beta {B_4}}.$ In
term of $\kappa $ the size of crossover region is defined $\mid \frac{%
T_0^{(2)}-T}{T_0^{(2)}}\mid <\frac 1\kappa $ in which the crossover occurs.
It is obvious that there is symmetry $\frac{T-T_0^{(2)}}{T_0^{(2)}}%
\rightarrow \frac{T_0^{(2)}-T}{T_0^{(2)}}$ in the crossover region $\mid 
\frac{T_0^{(2)}-T}{T_0^{(2)}}\mid <\frac 1\kappa .$

Away from the crossover region $\mid \frac{T_0^{(2)}-T}{T_0^{(2)}}\mid >%
\frac 1\kappa $, the tunneling rate reduces to the standard form 
\begin{equation}
Z_b=\frac 1\Delta \frac{\sqrt{\sum\limits_{n\neq 0}\omega _n^{(0)2}}}{\sqrt{%
\sum\limits_{n\neq 0,1}\omega _n^{(b)2}}}e^{\frac{-S_c}\hbar }  \label{2.18}
\end{equation}
where$\frac 1\Delta =\sqrt{\frac{S_c}{2\pi \hbar }}(\beta \hbar )$ which is
known by Faddeev-Popov technique. We can reach it only through normalizing
the eigenfunction of zero mode while now $N$ may be different from the upper
result for all modes being normalized $N=\frac{\sqrt{\omega _0^{(0)2}}\sqrt{%
\sum\limits_{n\neq 0}\omega _n^{(0)2}}}{2sinh(\frac{\beta \omega _0}2)}$.
The concrete parameter $\frac{\sqrt{\sum\limits_{n\neq 0}\omega _n^{(0)2}}}{%
\sqrt{\sum\limits_{n\neq 0,1}\omega _n^{(b)2}}}$ can be calculated only when 
$T\rightarrow T_0^{(2)}$\cite{27,31,32}.

\subsection{Universal law in crossover region}

Beyond steepest descent method, we have the formula eq. (\ref{2.4}) and eq. (%
\ref{2.5}), which is only needed in the crossover region 
\begin{equation}
\mid T-T_0^{(2)}\mid \leq T_0^{(2)}/\kappa  \label{2.19}
\end{equation}
where $\kappa =\frac{M\omega _R^2}2\sqrt{\frac \beta {B_4}}\gg 1.$ It has
been pointed out that there is a universal law in the crossover region of
second order transition\cite{10,11}. We use the following quantity to show
the universal law 
\begin{equation}
y=\Gamma \exp \left( \bigtriangleup U/k_BT\right) .  \label{2.20}
\end{equation}
which is a function of $\varepsilon $ but independent of the temperature $T$%
. According to the formula eq. (\ref{2.4}) and eq. (\ref{2.5}), we have the
universal law in the scaling region 
\begin{equation}
y/y_0=F\left( \xi /\xi _0\right)  \label{2.21}
\end{equation}
where $F\left( \xi \right) =%
%TCIMACRO{\limfunc{erfc}}
%BeginExpansion
\mathop{\rm erfc}
%EndExpansion
\left( \xi \right) \exp \left( \xi ^2\right) ,$ $\xi =T-T_0^{(2)}$, $\xi
_0=T_0^{(2)}/\kappa $ and $y_0=\frac{\omega _0}{2\pi }\frac{\left( \omega
_0^2+\omega _b^2\right) }2\sqrt{\frac{\beta \pi }{6c_4}}f_c\left[ \omega
_0,\omega _b\right] $.

\section{Decay rate in the crossover region at the boundary of second order
and first order transition}

$B_4=0$ is the boundary of the second order transition and the first order
one and the potential looks different from that of $B_4\neq 0$ : The
potentials changes slowly near the top and the bottom, but are rather steep
in the middle. Because there is divergence of $\kappa \rightarrow \infty $
at the point of $B_4=0$, the formula (\ref{2.12}) isn't available.

Above $T_0^{(2)},$ the decay process arises from sphelaron $x=x_b$, too. So
we have the same periodic path near the saddle point as formula (\ref{2.6}).
but the interactions terms of modes differ from eq. (\ref{2.11}).{\em \ }We
consider the sixth order terms of the two dangerous modes 
\begin{equation}
\Delta S=\beta \hbar \left[ 
\begin{array}{c}
\frac{M\omega _1^{\left( b\right) 2}}2(Y_1^2+Y_{-1}^2) \\ 
+B_4(Y_1^2+Y_{-1}^2)^2+B_6(Y_1^2+Y_{-1}^2)^3
\end{array}
\right]  \label{3.1}
\end{equation}
where $B_6=\frac 52c_6-\frac{2c_4^2}{M\omega _3^{\left( b\right) 2}}$. {\em %
Near the two-phase point not only the second term }$\frac 12\beta \hbar
M\sum\limits_{n=\pm 1}\omega _n^{\left( b\right) 2}Y_n^2${\em \ tends to
zero (Remember here }$\omega _{\pm 1}^2\rightarrow 0,${\em \ when
temperature turns to }$T_c${\em \ ) but also }$B_4${\em \ is a small
quantity. }We consider the sixth order terms of the two dangerous modes $%
B_6(Y_1^2+Y_{-1}^2)^3.\ $At the point $B_4=0$, there is no fourth term and
the formula is reduced to 
\begin{equation}
\frac 1\Lambda =\frac \beta 2\left[ \beta B_6\right] ^{-1/3}\int_0^\infty
dt\exp \left[ -\left( t^3-3\kappa ^{\prime }\varepsilon t\right) \right] .
\label{3.2}
\end{equation}
where $\kappa ^{\prime }=\frac{\beta M\omega _b^2}3\left[ \beta B_6\right]
^{-1/3}.$ Below $T_c,$ we transform $Y_n$ to $Y_n+X_n$ and have the same
form of $\frac 1\Lambda $.

The universal law in the crossover region $\mid T-T_c\mid \leq T_c/\kappa
^{\prime }$ is also defined as $y/y_0=F\left( \xi /\xi _0\right) $ where $%
F\left( \xi \right) =\int_0^\infty dt\exp \left[ -\left( t^3+3\xi t\right)
\right] ,$ $\xi =T-T_c$, $\xi _0=T_c/\kappa ^{\prime }$ and 
\begin{eqnarray}
y_0 &=&\frac{\omega _0}{12\omega _b}\left( \omega _0^2+\omega _b^2\right)
\label{3.3} \\
&&\left[ \beta B_6\right] ^{-1/3}\Gamma \left( \frac 13\right) f_c\left[
\omega _0,\omega _b\right]  \nonumber
\end{eqnarray}
, $\Gamma \left( x\right) $ is Gamma function.

\section{The effect of modes of the first order transition $B_4<0$}

The actual transition occur at the temperature when the two saddle points
have the same action. For $B_4>0$ the particles tunnel through the barrier
at the most {\it favorable} energy level $E(T)$ which goes down continuously
from the top of the barrier to the bottom of the potential with lowering
temperature. This corresponds to the second-order transition from thermal
activation to thermally assisted tunneling ( TAT) with no discontinuity of $%
d\Gamma /dt$ at $T_0$. And the transition temperature is given by $%
T_0^{(2)}. $ For the cases of $B_4<0,$ just below the top of the barrier is 
{\em unfavorable for tunneling}, the TAT mechanism is partially suppressed
and the first order transition occurs. Because the top of the barrier is
wider for $B_4<0$, the particle doesn't have more advantage to tunnel
through the barrier from the higher excited states than the lower ones.

Let us discuss the dangerous modes near the first order transition point$%
{\bf :}$ Before the eigenvalues of the two quasi-zero modes reach zero, the
crossover occurs and tunneling process is dominated by periodic instantons
just below the top of the barrier. In this case, there is really no
universal law. Above $T_c$ which is near $\omega _R/2\pi $, the sphelaron's
two dangerous modes may also play important role on the tunneling rate at
the transition point. Above the crossover temperature, the quasi-zero modes
of sphelaron have the same eigenfunction as other cases ($B_4\geq 0$ ) : $%
\sqrt{2}$ $\sin \frac{2\pi }L\tau $ and$\sqrt{2}\cos \frac{2\pi }L\tau $.
While there are many distinguished differences of the interaction of modes
between the case of $B_4>0$ and that of $B_4<0$ : The interaction between
modes is attractive of the lower order terms.

After integrating the amplitudes $Y_0$ and $Y_n$, we deal with the
quasi-zero modes to considerate the forth term

\begin{eqnarray}
\frac 1\Lambda &=&\frac \beta {2\pi }\int dY_1dY_{-1}  \label{4.1} \\
&&\ \times \exp -\beta \left[ \frac 12m\omega _1^{\left( b\right) 2}Y_1^2+%
\frac 12m\omega _{-1}^{\left( b\right) 2}Y_{-1}^2+B_4\left(
Y_1^2+Y_{-1}^2\right) ^2\right]  \nonumber \\
&=&\sqrt{\pi }\frac \kappa {2\omega _R^2}%
%TCIMACRO{\limfunc{erfi} }
%BeginExpansion
\mathop{\rm erfi}
%EndExpansion
\left( -\kappa \varepsilon \right) \exp \left( \kappa ^2\varepsilon ^2\right)
\nonumber
\end{eqnarray}
where $\kappa =\frac{M\omega _R^2}2\sqrt{\frac \beta {B_4}}$ and $%
\varepsilon =(1-\frac T{T_0^{(2)}})$ . The Quasi-zero modes give a divergent
contribution to the partition function which can be distorted into
complex-plane : $\int\limits_c^\infty \exp \left( x^2\right)
dx=\int\limits_0^\infty \exp \left( x^2\right) dx-\int\limits_0^c\exp \left(
x^2\right) dx=\frac{i\sqrt{\pi }}2-%
%TCIMACRO{\limfunc{erfi}}
%BeginExpansion
\mathop{\rm erfi}
%EndExpansion
\left( c\right) $. Here $%
%TCIMACRO{\limfunc{erfi}}
%BeginExpansion
\mathop{\rm erfi}
%EndExpansion
\left( c\right) =\int\limits_0^c\exp \left( x^2\right) dx.$ Because the real
part of the function $%
%TCIMACRO{\limfunc{erfi}}
%BeginExpansion
\mathop{\rm erfi}
%EndExpansion
\left( -\kappa \varepsilon \right) $ goes to zero as $\varepsilon
\rightarrow 0$, a dramatic phenomenon arises - first order transition may
suppress quantum tunneling rate (without considering higher order terms )!

Because the periodic trajectories of periodic instanton can't be written as
a Fourier series if they are far below the top of the barrier, the formula (%
\ref{4.1}) is useful only above the crossover temperature $T_0^{(1)}$.

Now we consider a {\em weak} {\em first order phase transition} which is
under the condition $T_0^{(1)}/T_0^{(2)}\leq 1-1/\kappa .$ For a weak first
order phase transition,{\em \ }the crossover temperature $T_0^{(1)}$ is in
the region $(T-T_0^{(2)})\leq T_0^{(2)}/\kappa $ which is shown in Fig. 3.
We may define {\em a remaining crossover region} above the crossover
temperature 
\begin{equation}
T_0^{(1)}\leq T\leq T_0^{(2)}+T_0^{(2)}/\kappa .  \label{4.2}
\end{equation}
We also use the quantity $y=\Gamma \exp \left( \bigtriangleup U/k_BT\right) $
to show the universal law as 
\begin{equation}
y/y_0=F(\xi /\xi _0)  \label{4.3}
\end{equation}
where $F(\xi )=%
%TCIMACRO{\limfunc{erfi}}
%BeginExpansion
\mathop{\rm erfi}
%EndExpansion
(\xi )\exp (\xi ^2)$ and $y_0=\frac{\omega _0}{2\pi }\frac{\left( \omega
_0^2+\omega _b^2\right) }2\sqrt{\frac{\beta \pi }{6c_4}}f_c\left[ \omega
_0,\omega _b\right] $. For the cases $T_0^{(1)}/T_0^{(2)}>1-1/\kappa ,$ the
first order transition is rather sharp, there is no crossover region at all.

\section{Scaling Law of Quantum-Classical Transition of the Escape Rate in Mn%
$_{12}$Ac}

A rather simple and experimentally important system which may exhibit the
first-order transition is the uniaxial spin model in a field parallel to $x$%
-axis $H_x$ described by the Hamiltonian 
\begin{equation}
{\cal H}=-DS_z^2-H_xS_x  \label{5.1}
\end{equation}
which is generic for problems of spin tunneling. This model is believed to
be a good approximation for the molecular magnet Mn$_{12}$Ac with $D$ the
anisotropy constant. Using the method of mapping the spin problem onto a
particle problem\cite{23,24}, we have the equivalent particle Hamiltonian as 
${\cal H}=\frac{p^2}{2m}+U(x),$ where 
\begin{equation}
U(x)=S^2D(h_x\cosh x-1)^2  \label{5.2}
\end{equation}
and $m=\frac 1{2D},$ $h_x=\frac{H_x}{2SD}$ and $S\gg 1.$ The minimum of the
effective potential, $x_0=$ $\cosh ^{-1}\frac 1{h_x},$ has been moved to
zero. Integrating the equation of motion with imaginary time variable one
obtains 
\begin{equation}
x_p(\tau )=2\tanh ^{-1}\left[ \tanh x_1%
%TCIMACRO{\limfunc{sn} }
%BeginExpansion
\mathop{\rm sn}
%EndExpansion
\left( \frac{\tau -\tau _0}{\xi _P},k\right) \right]  \label{5.3}
\end{equation}
\begin{equation}
\tanh ^2x_{1,2}=\frac{1-h_x\mp \sqrt{E^{\prime }}}{1+h_x\mp \sqrt{E^{\prime }%
}}  \nonumber
\end{equation}
where $%
%TCIMACRO{\limfunc{sn}}
%BeginExpansion
\mathop{\rm sn}
%EndExpansion
\left( \tau ,k\right) $ is the Jacobi elliptic function with modulus $k$ and
the complementary modulus $k^{\prime }=\sqrt{1-k^2}$, $E^{\prime }=\frac E{%
S^2D}.$ $\xi _P$ is the characteristic length of the periodic instanton
determined by the following equation 
\begin{equation}
\xi _P^2=\frac 1{S^2D^2}\left( 
\begin{array}{c}
1-h_x^2 \\ 
\times \left( \frac{(1+k^2)h_x^2-k^{\prime 2}(1-h_x^2)+h_x\sqrt{%
4h_x^2k^2+k^{\prime 4}}}{\left( 1+k^2\right) h_x^2+h_x\sqrt{%
4h_x^2k^2+k^{\prime 4}}}\right) ^2
\end{array}
\right) ^{-1}.  \label{5.4}
\end{equation}
This description corresponds to the movement of a pseudo-particle in the
inverted potential $-U(x)$ with energy $-E.$ The periodicity of the solution
(\ref{5.3}) is $\tau (E)=4\hbar \beta ,$ $\hbar \beta ={\rm K}(k)\xi _P,$
where ${\rm K}(k)$ is the complete elliptic integral of the first kind. The
Euclidean action of the periodic instanton configuration in the whole period
is 
\begin{equation}
S_p=2\int_{-\beta }^\beta \left[ m\stackrel{.}{x}_p^2\right] d\tau =W
\label{5.5}
\end{equation}
where 
\begin{eqnarray}
W &=&\frac 4{D\xi _P\alpha ^2}{\LARGE [}\left( \alpha ^4-k^2\right) {\rm \Pi 
}(\alpha ^2,k)+  \nonumber \\
&&k^2{\rm K}(k)+\alpha ^2\left( {\rm K}(k)-{\rm E}(k)\right) {\LARGE ]}
\end{eqnarray}
where $\alpha ^2=\tanh ^2x_1<k^2$ and ${\rm E}(k),$ ${\rm \Pi }(\alpha ^2,k)$
are the complete elliptic integral of second and third kind, respectively.

The period of oscillation $\tau (E)$ in the inverted potential $-U(x),$
i.e., the periodicity of the periodic instanton solution (\ref{5.3}), $\tau
(E)=4{\rm K}(k)\xi _P,$ can be equivalently calculated by 
\begin{equation}
\tau (E)=-\frac{dW(E)}{dE}=\sqrt{2m}\int_{-x_1}^{x_1}\frac{dx}{\sqrt{U(x)-E}}%
.  \label{5.6}
\end{equation}
Near the top of the barrier, $k\rightarrow 0,{\rm K}\rightarrow \pi /2,$ the
equation (\ref{5.6}) yields the previously known result $\tau =2\pi /\omega
_b$. Near the bottom, one has $k\rightarrow 1,$ and $\tau $ logarithmically
diverges. For $h_x<1/4,$ the dependence $\tau (E)$ is non-monotonic, and the
transition is first order.

Since both $S_0$ and $S_p$ are assumed to be large compared to $\hbar ,$ the
smaller of the two determines the actual escape rate. The calculation of the
temperature dependence of $S_{\min }$ is depicted in Fig.1 for $h_x=0.1$.
The solid line corresponds to the periodic instanton action $S_p$ while the
dashed one corresponds to the thermodynamic action $S_0.$ The actual
dependence of $S_{\min }(T)$ goes along the minimum of these two actions and
the first-order transition occurs at $T=T_0^{(1)}$ satisfying $%
T_2<T_0^{(1)}<T_1,$ where $k_BT_1=\hbar /\tau (E_1),$ $k_BT_2=\hbar /\tau
(\Delta U).$ The first derivative of $S_{\min }(T)$ is discontinuous at $%
T_0^{(1)},$ providing that the crossover from the thermal to the quantum
regime is the first order transition on temperature. Quite recently an
effective free energy $F=a\phi ^2+b\phi ^4+c\phi ^6$ for the transitions of
a spin system was introduced\cite{28} as in the Landau model of second order
phase transition. Here $a$ changes sign at the phase transition temperature $%
T=T_0^{(2)}=\frac{SD}{\pi k_B}\sqrt{h_x(1-h_x)}$ and $b=0$ corresponds to
the boundary between first- and second-order transitions. There indeed
exists a phase boundary between the first- and second-order transitions,
i.e., $h_x=\frac 14,$ at which the factor in front of $\phi ^2$ changes the
sign. At $h_x=0.3$ the minimum of $F$ remains $\Delta U$ for all $%
T>T_0^{(2)}.$ Below $T_0^{(2)}$ it continuously shifts from the top to the
bottom of the potential as temperature is lowered. This corresponds to the
second order transition from thermal activation to TAT. At $h_x=0.1$,
however, there can be one or two minima of $F,$ depending on the
temperature. The crossover between classical and quantum regimes occurs when
the two minima have the same free energy, which for $h_x=0.1$ takes place at 
$T_0^{(1)}=1.078T_0^{(2)}.$

Near the top of the barrier the potential has the form as

\begin{eqnarray}
U(x) &=&S^2D(h_x\cosh x-1)^2  \label{5.7} \\
&=&2S^2Dh_x-S^2Dh_x(h_x-1)x^2  \nonumber \\
&&\ +c_4x^4+c_6x^6...  \nonumber
\end{eqnarray}
where $c_4=\frac{S^2Dh_x(h_x-\frac 14)}3$ and $c_6=2S^2Dh_x(h_x-\frac 1{24}%
)/45.$ The potentials changes slowly near the top and the bottom, but are
rather steep in the middle for the cases $c_4<0$.

$h_x=1/4$ is the boundary of the second order transition and the first order
one. When $h_x>1/4$, we have the formula eq. (\ref{2.20}) and eq. (\ref{2.21}%
), which is only needed in the crossover region 
\begin{equation}
\mid T-T_0^{(2)}\mid \leq T_0^{(2)}/\kappa  \label{5.8}
\end{equation}
where 
\begin{equation}
\kappa =\left[ h_x(1-h_x)\right] ^{3/4}\sqrt{\frac{2\pi S}{h_x(h_x-\frac 14)}%
}\propto S^{1/2}\gg 1.  \label{5.9}
\end{equation}
It has been pointed out that there is a universal law in the crossover
region of second order transition\cite{11,12}. According to the formula eq. (%
\ref{2.20}) and eq. (\ref{2.21}), we have the universal law in the scaling
region 
\begin{equation}
y/y_0=F\left( \xi /\xi _0\right)  \label{5.10}
\end{equation}
where $F\left( \xi \right) =%
%TCIMACRO{\limfunc{erfc}}
%BeginExpansion
\mathop{\rm erfc}
%EndExpansion
\left( \xi \right) \exp \left( \xi ^2\right) ,$ $\xi =T-T_0^{(2)}$, $\xi
_0=T_0^{(2)}/\kappa $ and $y_0=\frac{\omega _0}{2\pi }\frac{\left( \omega
_0^2+\omega _b^2\right) }2\sqrt{\frac{\beta \pi }{6c_4}}f_c\left[ \omega
_0,\omega _b\right] $

For the case $h_x=1/4$, we have the crossover region $\mid T-T_0^{(2)}\mid
\leq T_0^{(2)}/\kappa ^{\prime }$ where 
\begin{equation}
\kappa ^{\prime }=6\left[ \frac{\pi h_x(1-h_x)S}9\right] ^{2/3}\left[
h_x(h_x-\frac 1{24})\right] ^{-1/3}\propto S^{2/3}\gg 1.  \label{5.11}
\end{equation}
>From the formula (\ref{3.3}), we have the universal law $y/y_0=F\left( \xi
/\xi _0\right) $ in the crossover region where $F\left( \xi \right)
=\int_0^\infty dt\exp \left[ -\left( t^3+3\xi t\right) \right] ,$ $\xi
=T-T_0^{(2)}$, $\xi _0=T_0^{(2)}/\kappa ^{\prime }$ and 
\begin{equation}
y_0=\frac{\omega _0\beta \left( \omega _0^2+\omega _b^2\right) }{24\pi }%
\left[ \frac 52\beta c_6\right] ^{-1/3}\Gamma \left( \frac 13\right)
f_c\left[ \omega _0,\omega _b\right] ,  \label{5.12}
\end{equation}
$\Gamma \left( x\right) $ is Gamma function which is shown in fig.2.

Now we consider a weak first order phase transition which is under the
condition $T_0^{(1)}/T_0^{(2)}\leq 1-\frac{\sqrt{\frac{h_x(h_x-\frac 14)}{%
2\pi S}}}{\left[ h_x(1-h_x)\right] ^{3/4}}.$ For a weak first order phase
transition,{\em \ }the crossover temperature $T_0^{(1)}$ is in the region $%
(T-T_0^{(2)})\leq T_0^{(2)}/\kappa $ which is shown in Fig. 3. We may define
a remaining crossover region above the crossover temperature 
\begin{equation}
T_0^{(1)}\leq T\leq T_0^{(2)}+T_0^{(2)}/\kappa .  \label{5.13}
\end{equation}
We also use the quantity $y=\Gamma \exp \left( \bigtriangleup U/k_BT\right) $
to show the universal law as $y/y_0=F(\xi /\xi _0)$ where $F(\xi )=%
%TCIMACRO{\limfunc{erfi}}
%BeginExpansion
\mathop{\rm erfi}
%EndExpansion
(\xi )\exp (\xi ^2)$ and $y_0=\frac{\omega _0}{2\pi }\frac{\left( \omega
_0^2+\omega _b^2\right) }2\sqrt{\frac{\beta \pi }{6c_4}}f_c\left[ \omega
_0,\omega _b\right] $. For the cases $T_0^{(1)}/T_0^{(2)}>1-\frac{\sqrt{%
\frac{h_x(h_x-\frac 14)}{2\pi S}}}{\left[ h_x(1-h_x)\right] ^{3/4}},$ the
first order transition is rather sharp, there is no crossover region at all.

\section{Conclusion}

In this paper we have shown that the decay rate $\Gamma $ can accurately be
determined near the crossover temperature in dissipative systems. Besides
considering the WKB exponential, we have calculated contribution of the
fluctuation modes around the saddle point and have given an extended account
of the previous study of crossover region\cite{10,11,12}. Near phase
transition point the fluctuation modes about the saddle points include two
dangerous modes whose contribution can't be calculated by the steepest
descent method and the higher order couplings are considered between the two
dangerous modes, near the point of 
\begin{equation}
B_4={\em \ }\frac 32c_4+\frac{9c_3^2}{2M\omega _b^2}-\frac{9c_3^2}{4M\omega
_2^{\left( b\right) 2}}=0  \label{6.1}
\end{equation}
sixth order need to be considered. The results can be easily used in Mn$%
_{12} $Ac of which the equation $B_4=\frac{S\left( S+1\right) Dh_x(h_x-\frac 
14)}2=0$ gives the phase boundary point $h_x=\frac 14$ in good agreement
with earlier works.

Another example is biaxial anisotropic ferromagnetic model ${\cal H}%
=K_1S_z^2+K_2S_y^2$ which describes XOY easy plane anisotropy and an easy
axis along the $x$ direction with the anisotropy constants $K_1>K_2>0$.
Mapped onto a particle problem, the equivalent particle Hamiltonian is 
\begin{equation}
{\cal H}=\frac 1{4K_1}\stackrel{.}{x}^2-K_2S(S+1)%
%TCIMACRO{\limfunc{sn} }
%BeginExpansion
\mathop{\rm sn}
%EndExpansion
{}^2\left( x,\lambda \right)  \label{6.2}
\end{equation}
where $%
%TCIMACRO{\limfunc{sn}}
%BeginExpansion
\mathop{\rm sn}
%EndExpansion
\left( \tau ,\lambda \right) $ is the Jacobi elliptic function with modulus $%
\lambda =K_2/K_1.$ From the equation $B_4=\frac{K_2S(S+1)\left( 1-\lambda
\right) \left( 1-2\lambda \right) }2=0,$ we obtains the phase boundary point 
$\lambda =\frac 12$ whichs confirm the results in ref. (\cite{30}).

The results of the application of a previous method\cite{7,8,9,10,12} is
developed for dealing with the quasi-zero modes and calculation of decay
rate in the crossover region which is beyond the steepest descent method.
The decay rate is valid for the entire interesting range of parameters in
the problem of MQT.

\begin{center}
{\bf Acknowledgment}
\end{center}

This work was supported by the National Natural Science Foundation of China.

Fig. 1

First-order transition from the thermal to the quantum region for Mn$_{12}$
molecular: $h_x=0.1.$

Fig. 2

Scale factor $F\left( \xi \right) $ at different sides of phase transition
point of $h_x=1/4.$

Fig. 3

The remaining scaling region of a weak first order phase transition is $%
T_0^{\left( 1\right) }\leq T\leq T_0^{\left( 2\right) }+T_0^{\left( 2\right)
}/\kappa .$

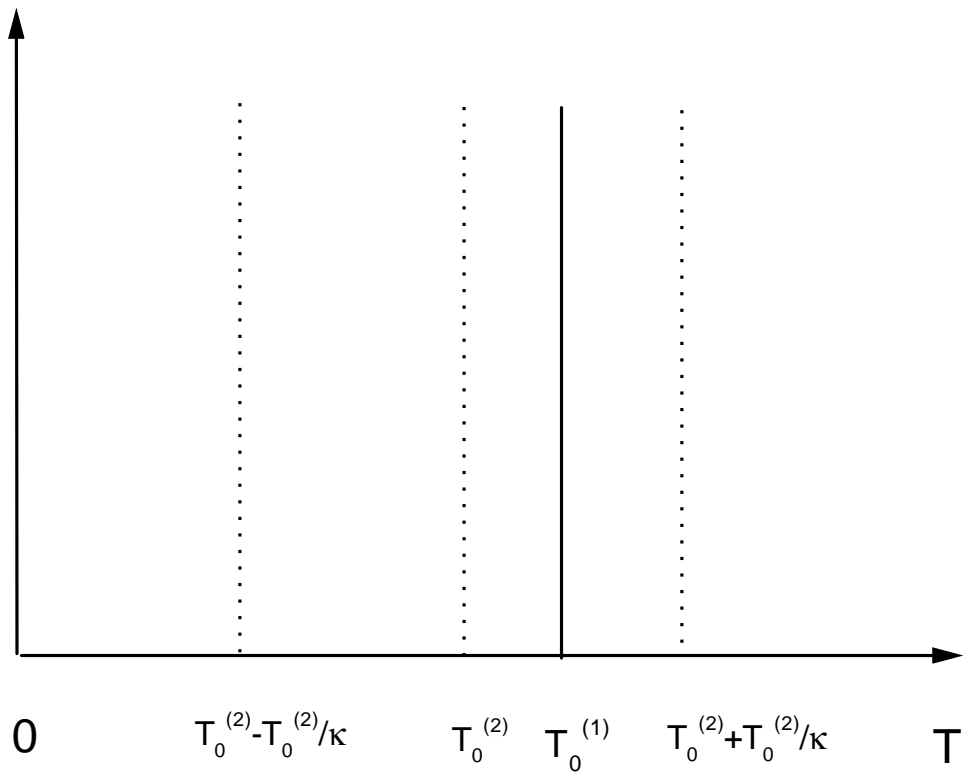
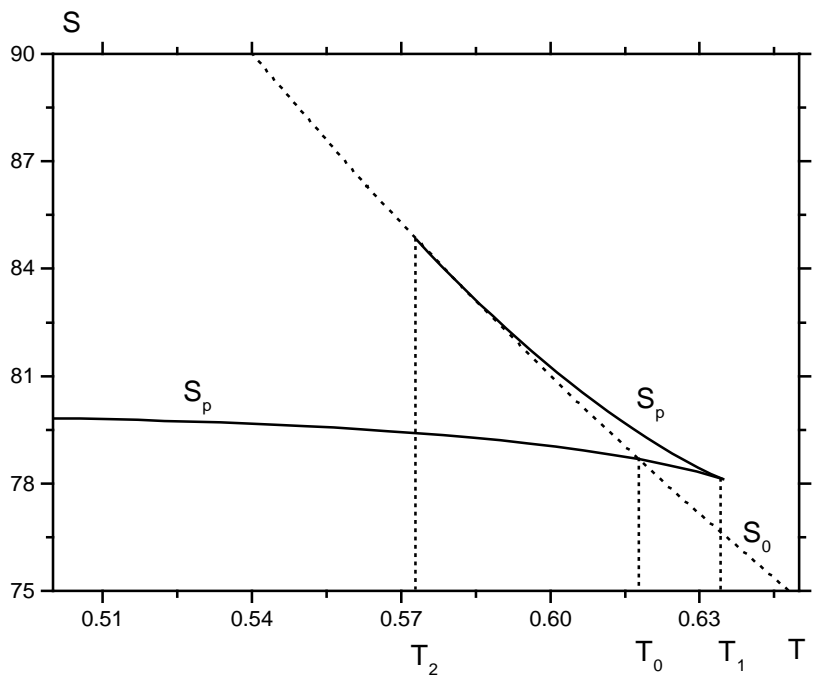
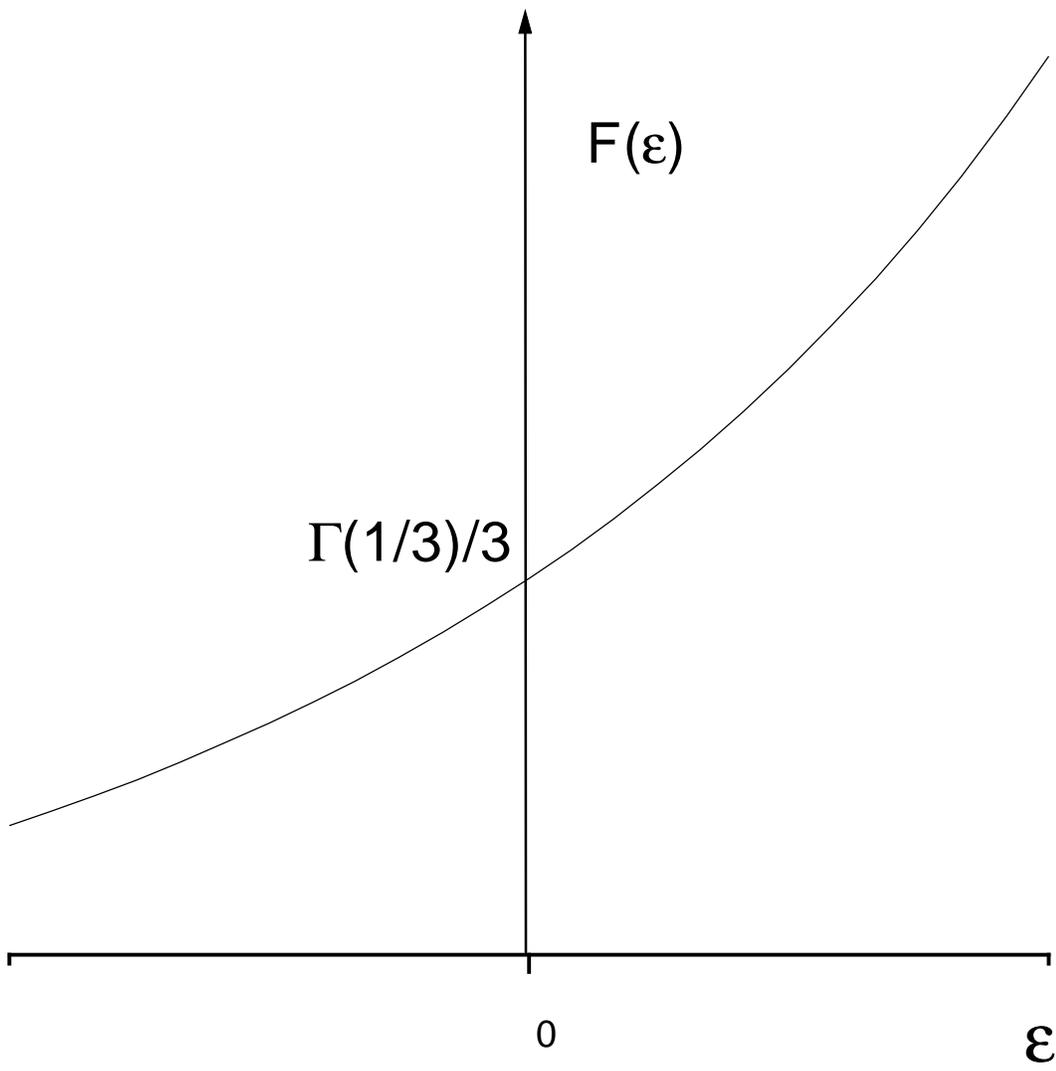
\end{document}